\documentclass[12pt]{article}
\usepackage[dvips]{graphicx,color}
\newfont{\Bbb}{msbm10 scaled 1200}
\newcommand{\mathbb}[1]{\mbox{\Bbb #1}}

\def\TL{\hfil$\displaystyle{##}$}
\def\TR{$\displaystyle{{}##}$\hfil}

\def\lbldef#1#2{\expandafter\gdef\csname #1\endcsname {#2}}
\def\eqn#1#2{\lbldef{#1}{(\ref{#1})}%
\begin{equation} #2 \label{#1} \end{equation}}
\def\eqalign#1{\vcenter{\openup1\jot
    \halign{\strut\span\TL & \span\TR\cr #1 \cr
   }}}

\def\href#1#2{#2}

\def\dplus{=\hskip-5pt \raise 0.7pt\hbox{${}_\vert$}
^{\phantom 7}}
\def\dplusup{=\hskip-5.1pt \raise 5.4pt\hbox{${}_\vert$}
^{\phantom 7}}
\def\dplus{=\hskip-4.8pt \raise 0.7pt\hbox{${}_\vert$}
^{\phantom 7}}
\font\small=cmr8

\def\pmb#1{\setbox0=\hbox{#1} \kern-.025em\copy0\kern-\wd0
\kern0.05em\copy0\kern-\wd0 \kern-.025em\raise.0433em\box0}

\def\ubx{{\underline{x}}}
\def\uby{{\underline{y}}}

\font\mybb=msbm10 at 11pt

\def\bb#1{\hbox{\mybb#1}}

\def\bZ {\bb{Z}}
\def\bR {\bb{R}}
\def\cN{{\cal{N}}}
\def\rre {{\rm Re}}
\def\iim {{\rm Im}}

\def\e  {\epsilon}

\font\mybb=msbm10 at 12pt
\def\bb#1{\hbox{\mybb#1}}

\def\C{\mkern1mu\raise2.2pt\hbox{$\scriptscriptstyle|$}
\mkern-7mu{\rm C}}

\def\log{{\rm log}}

\def\cP{{\cal{P}}}

\def\l {\lambda}
\def\d{\delta}

\def\a{\alpha}

\def\t{\tau}
\def\m{\mu}

\def\g{\gamma}
\def\pd{\partial_}

\def\g{\gamma}
\def\k{\kappa}
\def\f{\phi}
\def\s{\sigma}
\def\e{\epsilon}

\def\T{\Theta}

\def\x {\chi}

\newcommand{\beq}{\begin{equation}}
\newcommand{\eeq}{\end{equation}}
\newcommand{\ber}{\begin{eqnarray}}
\newcommand{\eer}{\end{eqnarray}}

\newcommand{\beqar}{\begin{eqnarray}}

\newcommand{\eeqar}{\end{eqnarray}}

\newcommand{\ba}{\begin{eqnarray}}
\newcommand{\ea}{\end{eqnarray}}

\newcommand{\dsl}
  {\kern.06em\hbox{\raise.15ex\hbox{$/$}\kern-.56em\hbox{$\partial$}}}

\newcommand{\eeqarr}{\end{eqnarray}}
\newcommand{\ZZ}{{\rm \kern 0.275em Z \kern -0.92em Z}\;}

\begin{document}
\baselineskip=15.5pt
\pagestyle{plain}
\setcounter{page}{1}
\begin{titlepage}

\vskip -.8cm

\rightline{\small{\tt September 2001}}

\begin{center}

\vskip 1.7 cm

{\LARGE {Stringy Domain Walls of $\cN=1$, $D=4$ Supergravity}}

\vskip 1.5cm
{\large
J. Gutowski
}
\vskip 1.2cm

Department of Physics,  Queen Mary College,
Mile End, London, E1 4NS
\vskip 0.5cm

\vspace{1cm}

{\bf Abstract}

\end{center}

We examine domain wall solutions of $\cN=1$, $D=4$ supergravity
which preserve half of the supersymmetry and
arise from Euclidean M2-brane instantons on M5-branes
wrapping associative 3-cycles of $G_2$-holonomy manifolds.
We also investigate composite
solutions which break an additional half of the supersymmetry.

\noindent

\end{titlepage}

\newpage


\section{Introduction}

There has been considerable interest shown in recent months in the
geometry of seven-dimensional manifolds with holonomy $G_2$.
At the level of ten and eleven dimensional supergravity theories,
the fact that such manifolds preserve a proportion of supersymmetry
and possess parallel spinors means that new solutions corresponding
to branes in non-trivial backgrounds may be constructed. In this context,
new non-compact $G_2$ metrics are particularly important \cite{giba},
\cite{gibb}, \cite{gomis}; and they are related to various types
of domain wall geometries \cite{gibnew}. Furthermore,
such solutions may be used to probe possible dualities between
 supergravity
theories and field theory \cite{duala}, \cite{dualb},
\cite{dualbb}, \cite{dualc}. Another important aspect of $G_2$
holonomy manifolds is the fact that eleven dimensional
supergravity compactified on a 7-manifold of $G_2$ holonomy
reduces to $\cN=1$, $D=4$ supergravity, which is of special
significance phenomenologically.

    In this paper we shall examine some aspects of supersymmetric
solutions of a $\cN=1$, $D=4$ supergravity theory obtained from
 such a $G_2$
compactification. This investigation is a continuation of that
presented in
\cite{papa}, in which some aspects of the moduli space of $G_2$
structures
examined in \cite{joyce} and \cite{hitchin} were used to simplify
the couplings of the
four-dimensional theory, and some solutions were presented.
The examples we consider are motivated by M-brane configurations
in eleven dimensions wrapped in various fashions on a compact $G_2$
 holonomy
manifold. We
shall concentrate on solutions which are associated with the wrapping
of $M5$-branes on supersymmetric cycles of the $G_2$ manifold
 associated with calibrated geometries.
For example, the geometry of a $M5$-brane in directions
012389 wrapping an associative supersymmetric
3-cycle of the $G_2$ manifold along 1234567 may be represented
 schematically as
\eqn{assocwrp}{
\eqalign{
M5 & :  0 \  \ , \ 1 \ \ , \ 2 \ \ , \ 3 \ \ \  , \ * \  \ , \ * \ \
, \ * \  \ , \ * \ \ , \ 8 \ \ , \ 9 \ \ , \ *
\cr
G_2 & :  - \ , \ X \ , \ X \ , \ X \ , \ X \ , \ X \ , \ X \ , \ X \
, \ - \ , \ - \ , \ - \ ,}}
and compactifying on the $G_2$ manifold produces a 2-brane solution,
i.e. a domain wall.
We remark that such solutions
do not possess, a priori from these constraints, the physical
properties of domain
walls summarized in \cite{supdw}, such as disconnected supersymmetric
 extrema of the potential.
This depends on the form of the superpotential used in the theory,
 and we
shall examine this in more detail later.

It is also possible for an $M5$-brane to wrap a co-associative
 supersymmetric
calibrated 4-cycle of the $G_2$ manifold. Using the same notation
 this is
given by
\eqn{coassocwrp}{
\eqalign{
M5 & :  0 \ \ , \ 1 \ \ , \ * \ \ , \ 3 \ \  , \ 4 \ \ \ , \ * \ \
, \ 6 \  \ , \ * \ \ , \ 8 \ \ , \ * \ \ , \ *
\cr
G_2 & :  - \ , \ X \ , \ X \ , \ X \ , \ X \ , \ X \ , \ X \ , \ X \
, \ - \ , \ - \ , \ - \ .}}
This solution corresponds to a stringy cosmic string solution in
four dimensions
\cite{comstr}, \cite{cosstr}.
Both \assocwrp\ and \coassocwrp\ preserve $1 \over 16$ of the
 $D=11$ supersymmetry,
i.e. $1 \over 2$ of the $D=4$ supersymmetry. It is also possible
 to take the
orthogonal intersection of two $M5$-branes over a 3-brane,
 wrapping one of the
$M5$-branes on a co-associative cycle, and the other on an
associative cycle.
This gives
\eqn{ascoassocwrp}{
\eqalign{
M5 & :  0 \  \ , \ 1 \ \ , \ 2 \ \ , \ 3 \ \ \  , \ * \  \ , \ * \ \
, \ * \  \ , \ * \ \ , \ 8 \ \ , \ 9 \ \ , \ *
\cr
M5 & :  0 \ \ , \ 1 \ \ , \ * \ \ , \ 3 \ \  , \ 4 \ \ \ , \ * \ \
, \ 6 \  \ , \ * \ \ , \ 8 \ \ , \ * \ \ , \ *
\cr
G_2 & :  - \ , \ X \ , \ X \ , \ X \ , \ X \ , \ X \ , \ X \ , \ X \
, \ - \ , \ - \ , \ - \ .}}
This solution breaks another half of the eleven dimensional
 supersymmetry, and
preserves only $1 \over 4$ of the $D=4$ supersymmetry. It is a
 composite string and
domain wall solution.

The plan of this paper is as follows. In section 2 we present the
truncated $\cN=1$, $D=4$ supergravity action together with its
field equations and supersymmetry constraints. We also summarize
how geometric constraints imposed by requiring that the
 compactifying
7-manifold has $G_2$ holonomy simplify the couplings,
 and discuss some
properties of superpotentials. In section 3 we derive various
 supersymmetry
constraints associated with domain wall and composite string and
domain wall solutions preserving $1 \over 2$ and $1 \over 4$ of
 the $\cN=1$,
$D=4$ supersymmetry respectively. In section 4 we investigate
 numerically
the properties of domain wall type solutions obtained from various
M2-brane instanton superpotentials. In section 5 we present
 some conclusions.

\section{$N=1$ $D=4$ Supergravity}

In this section we summarize some important details of
 $\cN=1$, $D=4$ supergravity. We also present the
constraints obtained on the various couplings when the
four dimensional theory is obtained from compactification of
eleven-dimensional supergravity on a manifold of $G_2$ holonomy.

\subsection{Supergravity Action and Killing Spinor Equations}

The geometric data that determine the
various couplings of the $\cN=1$ four-dimensional
supergravity theory consists of
$n$ vector and
$m$ chiral multiplets together with

\begin{itemize}

\item{(i)} A K\"ahler-Hodge manifold $M$ of
complex dimension $m$ with K\"ahler
potential $K$.

\item{(ii)} A  vector bundle $E$ over $M$ of
rank $n$ for which its complexified  symmetric product
admits a holomorphic section $h$.

\item{(iii)} A locally defined holomorphic function $f$ on $M$.

\item{(iv)} Sigma model maps, $z$, from the four-dimensional
spacetime $\Sigma$ into the manifold $M$.

\item{(v)} A principal bundle $P$ on the
four-dimensional spacetime
$\Sigma$ with fibre the abelian group $U(1)^n$ such that
the pull back of $E$ with respect to $z$ is isomorphic to
$P\times_{U(1)^n} {\cal L}U(1)^n$, where ${\cal L}U(1)^n$
is the Lie algebra of $U(1)^n$.

\end{itemize}

The bosonic part  of the $\cN=1, D=4$
supergravity action is \cite{cremact}, \cite{wittact}, \cite{baggact}
\eqn{sugract}{
L=\sqrt{-g}\big[ {1\over2} R(g)-
{1\over4} {\rm Re}h_{ab} F^a_{MN} F^b{}^{MN} +{1\over4}
{\rm Im}h_{ab} F^a_{MN} {}^\star F^b{}^{MN}-
\g_{i\bar j}\partial_M z^i\partial^M z^{\bar
j}-V \big]}
where
\eqn{poteq}{
V=e^K[\g^{i\bar j} D_if  D_{\bar j}{\bar f}- 3 |f|^2]
+{1\over2} D_a D^a\ ,}
\eqn{gageq}{
F^a_{MN}=\partial_M A_N^a-\partial_N A_M^a\ ,}
\eqn{coveq}{
D_i f=\partial_i f+\partial_i K f\ ,}
$A_N^a$ are $U(1)$ (Maxwell) gauge potentials and the
$D_a$ are constants associated to a Fayet-Iliopoulos term.
The gauge indices $a, b=1, \dots, n$ are
raised
and lowered with ${\rm Re} h_{ab}$; $i,j=1, \dots, m$ and
$M,N=0,\dots,3$ are holomorphic sigma model manifold
and spacetime indices, respectively.

In this paper, we shall consider solutions of $\cN=1$, $D=4$
supergravity which preserve some proportion of the supersymmetry.
The Killing spinor equations of \sugract\ in a bosonic background
are
most  conveniently expressed in terms
of a real 4-component Majorana spinor $\e$ as
\eqn{supa}{
2 \big( \pd{M} +{1 \over 4} \omega_{M
{\underline A} {\underline B}}
\Gamma^{{\underline A} {\underline B}} \big) \e
-\big(  {\rm Im} (K_i \partial_M z^i)
+e^{K \over 2} \big( {\rm Re} f - {\rm Im} f \Gamma^5 \big)
\Gamma_M \e=0\ ,}
\eqn{supb}{
\big( -{1 \over 2} F{}^a{}_{MN} \Gamma^{MN} +
\Gamma^5 D^a \big) \e=0}
and
\eqn{supc}{\eqalign{
\big( {\rm Re} (\partial_M z^i) -
\Gamma^5 {\rm Im} (\partial_M z^i) \big) \Gamma^M \e
-e^{K \over 2} \big( {\rm Re} (\g^{i {\bar j}}
D_{\bar j} {\bar f})
- \Gamma^5
{\rm Im} (\g^{i {\bar j}} D_{\bar j} {\bar f}) \big)
\e =0\ ,}}
where underlined indices ${\underline A}, \ {\underline B}$
 denote tangent frame
indices and $\Gamma^5 =
 \Gamma^{\underline{0}} \Gamma^{\underline{1}}
\Gamma^{\underline{2}}
\Gamma^{\underline{3}}$. For our spinor conventions
see the appendix.

The field equations of the supergravity action \sugract\
 are the following:

\begin{itemize}

\item{(1)} The Einstein equations are:
\eqn{einga}{
\eqalign{
G_{MN} - \rre h_{ab} F{}^a{}_{ML} F{}^b{}_N{}^L
- 2\g_{i \bar j} \partial_{(M} z^i \partial_{N)} z^{\bar j}
\cr
+g_{MN} \big({1 \over 4} \rre h_{ab} F{}^a{}_{LP} F^{bLP}
+ \g_{i \bar j}
 \partial_L z^i \partial^L z^{\bar j} +V \big) =0\ .}}

\item{(2)} The  Maxwell field equations  are:

\eqn{maxgb}{
\pd{M} \big[ \sqrt{-g} \big( \rre h_{ab} F^{bMN} -
 \iim h_{ab} {}^\star F^{b MN} \big) \big]\ .}

\item{(3)} The scalar equations; varying $z^\ell$
gives the equation
\eqn{scalgc}{
\eqalign{
-{1 \over 8} \pd{\ell} h_{ab} F{}^a{}_{MN} F^{b MN} -
\pd{\ell}V - {i \over 8} \pd{\ell} h_{ab} F{}^a{}_{MN}
{}^\star F^{b MN}
\cr
+\g_{\ell \bar j}
\big( \nabla_M \partial^M z^{\bar j}
+ \Gamma{}^{\bar j}{}_{\bar i \bar k} \pd{M} z^{\bar i}
\partial^M z^{\bar k} \big)=0\ ,}}
where $\nabla_M$ is the covariant derivative with
respect to the Levi-Civita
connection of the spacetime metric and
\eqn{varipot}{
\pd{\ell} V = \pd{\ell} \big( e^K \g^{i \bar j} D_i f\big)
D_{\bar j} {\bar f}  -2 e^K {\bar f} D_\ell f
+{1 \over 2} \pd{\ell} (D_a D^a)\ .}
Taking the conjugate of this equation, one
obtains the field equation for
$z^{\bar \ell}$.

\end{itemize}

We remark that stringy cosmic string solutions with non-vanishing
 Fayet-Iliopoulos
terms arising from taking $D^a \neq 0$ have been found in
\cite{gutpap}. For the remainder
of this paper we shall set $D^a=0$.

\subsection{M-Theory Compactification on $G_2$ Manifolds}

The relationship between the supergravity action \sugract\ and the
action of eleven dimensional supergravity compactified on manifolds of
$G_2$ holonomy has been examined in detail in \cite{paptown}
 and \cite{papa}. We shall summarize
some of the results which are of particular use for our
 purposes. Suppose
that the $G_2$-holonomy manifold is $N$;
$\{\phi_i; i=1, \dots, m=b_3\}$
is a basis of $H^3(N, \bR)$ and
 $\{\omega_a; a=1, \dots, n=b_2\}$ is a basis of $H^2(N, \bR)$. Then
the complex sigma model co-ordinates may be written as
$z^i = -{1 \over 2} (s^i +i p^i)$ for $s^i$, $p^i \in \bR$.
Setting
\eqn{calib}{
\f = s^i \f_i}
the various couplings of the four dimensional theory obtained from
the compactification of eleven dimensional supergravity are
\eqn{coupl}{
\eqalign{
ds^2&=\gamma_{i\bar j}dz^i d\bar z^j=k_{ij}(s) ds^i ds^j
+m_{ij}(s) dp^i dp^j
\cr
m_{ij}(s)&={1\over 4 \int_N \sqrt{G}\, d^7y} \int_N\,
\sqrt{G}\, d^7y\, (\phi_i, \phi_j)
\cr
{\rm Re} h_{ab}(s)&={1\over2}\int_N \sqrt{G}\,
d^7y\,(\omega_a, \omega_b)={1\over2}\int
\omega_a\wedge *\omega_b=-{1\over2}\int_N
\omega_a\wedge
\omega_b\wedge \phi
\cr
{\rm Im} h_{ab}(p)&=-{1\over2} p^i
 \int_N \omega_a\wedge\omega_b\wedge \phi_i=
-{1\over2} p^i C_{iab}\ .}}
We shall denote the volume of the compact $G_2$ holonomy manifold by
$\T$. With respect to the $G_2$ moduli co-ordinates described here,
$\T= \T( s^i)$ is homogeneous of degree $7 \over 3$ with respect
 to the $s^i$,
and the K\"ahler potential is related to $\T$ by
\eqn{kahp}{
K =-{3 \over 7} \log \ \T \ .}
It remains to consider the role played by the superpotential $f$. Such
terms do not arise from direct compactifications of eleven dimensional
supergravity using the ansatz
presented above to four dimensions. However, a superpotential
 may originate from
some non-vanishing 4-form flux $F^0$  along the compact
 directions. Such a superpotential
has been considered in \cite{guk}, \cite{her} and \cite{ach}.
In this case, the potential is specified via
\eqn{suppa}{
\rre \ f (z) = \int_N \f \wedge F^0 \ .}
However, it has been argued that obtaining $f$ from the
 4-form flux is not
consistent with the compactness of the $G_2$ manifold \cite{beck}.
An alternative mechanism for generating potentials is
the wrapping of M2-branes on associative 3-cycles of $G_2$ manifolds.
The contribution to the superpotential from such an M2-brane instanton
is \cite{moore}
\eqn{supp}{
\triangle f (z) = \m e^{z}}
where $\m>0$ is constant and $z =z^i \d_i$ for real constants $\d_i$.
It has however been argued that there generically exist
 obstructions to the
construction of a locally smooth moduli space of associative cycles
\cite{macl}, although there are special cases when there does exist
a smooth moduli space. More recently, it has been
proposed in \cite{oog} and \cite{moore} that the contribution
 from multiple M2-brane instantons
may be obtained by taking the sum
\eqn{sumsup}{
f(z) = \m \sum_{n=1}^\infty {e^{n z} \over n^2}\ .}
We shall concentrate on supergravity solutions corresponding
to \supp\ and
\sumsup\ .

\section{Stringy Domain Wall Solutions}

In order to investigate domain wall solutions,
and composite string and domain wall solutions
we shall consider the following ansatz;
\eqn{apd}{
\eqalign{
ds^2&=A^2( w, \bar w) ds^2(\bR^{1,1})+ ds^2_{(2)}
\cr
z^i&=z^i(w, \bar w)
\cr
A^a&=0}}
where $ds^2_{(2)}$ is a metric on the manifold
spanned by $w, \bar w$
where $w=x+iy$ and ${\bar w} = x-iy$ for $x, \ y \in \bR$.
Without loss of generality, we shall take
\eqn{ape}{
ds^2_{(2)} = B^2(x,y) (dx^2+dy^2)}
to be diagonal, using the fact that any metric
on a Riemann surface is locally conformally flat.

Substituting this ansatz into the Killing spinor equations,
we find that
\eqn{susya}{
\pd{x}A \Gamma_\ubx + \pd{y} A \Gamma_\uby \e
+ ABe^{K \over 2} \big( {\rm Re}f + {\rm Im}f \Gamma^5
\big) \e =0}
together with
\eqn{susyb}{
\eqalign{
2 \pd{x} \e - \pd{x} \log A \e + \pd{y} \log {B \over A}
\Gamma_\ubx \Gamma_\uby \e
- \Gamma^5 {\rm Im} (K_i \pd{x} z^i) \e & =0
\cr
2 \pd{y} \e - \pd{y} \log A \e - \pd{x} \log {B \over A}
\Gamma_\ubx \Gamma_\uby \e
- \Gamma^5 {\rm Im} (K_i \pd{y} z^i) \e & =0}}
and
\eqn{susyc}{
\eqalign{
\big( \rre \pd{x} z^i - \Gamma^5 \iim
\pd{x} z^i \big) \Gamma_\ubx \e
+ \big( \rre \pd{y} z^i - \Gamma^5 \iim
\pd{y} z^i \big) \Gamma_\uby \e
\cr
- Be^{K \over 2} \big( \rre (\g^{i \bar j}
D_{\bar j} {\bar f}) - \Gamma^5 \iim
 (\g^{i \bar j} D_{\bar j} {\bar f}) \big) \e =0}}

The solutions which we shall concentrate on have $f \neq 0$
and preserve $1 \over 4$ of the supersymmetry. We may begin by
examining \susyc\ . If we work in a real basis, so that
\eqn{aph}{
\e = \pmatrix{ \e_1 \cr \e_2}}
with $\e_1$, $\e_2$ real; then \susyc\  implies
\eqn{api}{
\eqalign{
(\s^1 - 1) \big[ 2i \pd{w} z^i {\bar{\eta}}
+Be^{K \over 2} \g^{i \bar j} D_{\bar j} {\bar f}
\eta \big] & =0
\cr
(\s^1 +1) \big[ -2i \pd{\bar w} z^i {\bar \eta}
+Be^{K \over 2} \g^{i \bar j} D_{\bar j} {\bar f}
\eta \big] & =0}}
where $\eta = \e_1 +i \e_2$. Suppose now
that there exists $i$ such that
$\g^{i \bar j} D_{\bar j} {\bar f}=0$. Then for these $i$,
these equations may be solved by taking $z^i$ constant.
Alternatively, one may have $z^i = z^i (w)$
non-constant holomorphic with $\Gamma^5
\Gamma_{\ubx} \Gamma_{\uby} \e = - \e$;
or $z^i = z^i ({\bar w})$  non-constant anti-holomorphic
with $\Gamma^5 \Gamma_{\ubx} \Gamma_{\uby} \e =  \e$
(however if there exists more
that one value of $i$ such that
$\g^{i \bar j} D_{\bar j} {\bar f}=0$ then one cannot
have a supersymmetric solution with a mixture of
corresponding non-constant holomorphic
and anti-holomorphic complex scalars).

 Suppose now we consider $i$ for which
$\g^{i \bar j} D_{\bar j} {\bar f} \neq 0$. Define
\eqn{apj}{
\eqalign{
\psi^i & = 2i \pd{w} z^i \big[ Be^{K \over 2}
\g^{i \bar j} D_{\bar j} {\bar f} \big]^{-1}
\cr
\t^i & = -2i \pd{\bar w} z^i \big[ Be^{K \over 2}
\g^{i \bar j} D_{\bar j} {\bar f} \big]^{-1}}}
Then one requires for these $i$;
\eqn{apk}{
\eqalign{
(1-\s^1) \big(\psi^i {\bar \eta} + \eta \big) & =0
\cr
(1+ \s^1) \big( \t^i {\bar \eta} + \eta \big) & =0}}
There are several possibilities. Firstly, note that
one cannot have a supersymmetric solution with both
 $\psi^i=\t^i=0$. If $\psi^i=0$ then it turns out that
$\Gamma^5 \Gamma_{\ubx} \Gamma_{\uby} \e =  -\e$.
If $\t^i=0$, however, then
$\Gamma^5 \Gamma_{\ubx} \Gamma_{\uby} \e =  \e$.
Alternatively, one may have $\psi^i$, $\t^i$ both
nonzero. It turns out that if both
$|\psi^i| \neq 1$ and $|\t^i| \neq 1$ then the
 solution cannot be supersymmetric.
If however, $|\psi^i| \neq 1$ but $|\t^i|=1$ then one has
$\Gamma^5 \Gamma_{\ubx} \Gamma_{\uby} \e =  -\e$. Another
possibility is to take $|\t^i| \neq 1$ and
 $|\psi^i|=1$; then $\Gamma^5 \Gamma_{\ubx}
\Gamma_{\uby} \e =  \e$. We shall see that these solutions
generically preserve $1 \over 4$ of the supersymmetry. They
correspond to a superposition of string and domain wall solutions.
Before considering this case, we shall consider the remaining
case in which we take  $|\t^i|=|\psi^i|=1$. These
conditions have been
examined previously in \cite{papa} and they give solutions
which preserve
half of the supersymmetry. It is useful to recap these results here.

\subsection{Half Supersymmetric Solutions}

Writing $\psi^i =
e^{i \theta^i}$, $\t^i = e^{i \f^i}$ for real
$\theta^i$, $\f^i$,
the supersymmetry constraint \susyc\ for $|\t^i| = 1$
 and $|\psi^i|=1$  is satisfied by taking
\eqn{weird}{
\eqalign{
\e_1 & = \sin \f^i \pmatrix{ \l^i \cr \l^i} + \sin
\theta^i \pmatrix{- \m^i \cr \m^i}
\cr
\e_2 & =  -(1+ \cos \f^i) \pmatrix{ \l^i \cr \l^i} -
(1+ \cos \theta^i)  \pmatrix{- \m^i \cr \m^i}}}
for real $\m^i$, $\l^i$.
Analogous reasoning may be used to consider \susya\ .
 In particular, \susya\ may be written as
\eqn{aplxx}{
\eqalign{
(\s^1 -1) \big[ 2i \pd{w} A {\bar \eta} -AB
e^{K \over 2} {\bar f} \eta \big] & =0
\cr
(\s^1 +1) \big[ -2i \pd{\bar w}A {\bar \eta} -AB
e^{K \over 2} {\bar f} \eta \big] & =0 \ .}}
Defining
\eqn{apmxx}{
\eqalign{
\Omega & = -2i \pd{w} A \big(ABe^{K \over 2}
{\bar f} \big)^{-1}
\cr
\Lambda & = 2i \pd{\bar w} A  \big(ABe^{K \over 2}
 {\bar f} \big)^{-1}}}
we note that \susya\ is equivalent to
\eqn{apnxx}{
\eqalign{
(\s^1 -1) (\Omega {\bar \eta} + \eta) & =0
\cr
(\s^1 +1) (\Lambda {\bar \eta} + \eta) & =0} \ .}
Hence the reasoning used to determine the various
possible values of $\psi^i$, $\t^i$
also applies to $\Omega$ and $\Lambda$.

So, we have shown that
\susya\ and \susyc\ imply that
 $\Gamma^5 \Gamma_{\ubx} \Gamma_{\uby} \e \neq \pm \e$.
Furthermore, if $\g^{i \bar j} D_{\bar j} {\bar f}=0$
then $z^i$ is constant, and if $\g^{i \bar j}
D_{\bar j} {\bar f} \neq 0$ then
$\psi^i = \Omega$ and $\t^i = \Lambda$ with
$|\Omega|=|\Lambda|=1$. $\e$ is given by \weird\ .

It is also necessary to examine \susyb\ .
This constraint may be rewritten as
\eqn{appxx}{
4 \pd{\bar w} {\hat \e} -2i  \pd{\bar w} \log
{B \over A} \Gamma_\ubx \Gamma_\uby {\hat \e}
+i \Gamma^5 (-\pd{\bar w}K +2 K_i \pd{\bar w} z^i)
{\hat \e}=0 \ ,}
where ${\hat \e} = A^{-{1 \over 2}} \e$.
In this case $\t^i = \Lambda$ and $\psi^i = \Omega$
 imply (for $f \neq 0$)
\eqn{apqxx}{
\eqalign{
- \pd{\bar w} z^i & = A^{-1} \pd{\bar w} A {\bar f}^{-1}
 \g^{i \bar j} D_{\bar j} {\bar f}
\cr
- \pd{ w} z^i & = A^{-1} \pd{ w} A {\bar f}^{-1}
\g^{i \bar j} D_{\bar j} {\bar f}}}
and we solve the supersymmetry constraints
by taking $\Lambda= e^{i \f}$, $\Omega= e^{i \theta}$,
for $\theta$, $\f \in \bR$ with
\eqn{aprxx}{
\eqalign{
{\hat \e}_1 &  = \sin \f \pmatrix{{\hat \l} \cr
{\hat \l}} + \sin \theta \pmatrix{-{\hat \m} \cr {\hat \m}}
\cr
{\hat \e}_2 & = -(1+\cos \f) \pmatrix{{\hat \l} \cr {\hat\l}}
-(1+\cos \theta) \pmatrix{-{\hat \m} \cr {\hat \m}}}}
where ${\hat \l}$, ${\hat \m} \in \bR$. Then \susyb\
implies that
\eqn{apsxx}{
\eqalign{
4 \pd{\bar w} ({\hat \l} \sin \f) +i (1+\cos \f)
(\pd{\bar w}
(K+2 \log {B \over A})-2 K_i \pd{\bar w} z^i) {\hat \l} & =0
\cr
-4 \pd{\bar w}((1+\cos \f) {\hat \l})
+i \sin \f (\pd{\bar w} (K+2 \log {B \over A})-
2 K_i \pd{\bar w} z^i) {\hat \l} & =0
\cr
4 \pd{\bar w} ({\hat \m} \sin \theta)
+i (1+\cos \theta) (\pd{\bar w} (K-2 \log {B \over A})-
2 K_i \pd{\bar w} z^i) {\hat \m} & =0
\cr
-4 \pd{\bar w}((1+\cos \theta) {\hat \m})
+i \sin \theta (\pd{\bar w} (K-2 \log {B \over A})-
2 K_i \pd{\bar w} z^i) {\hat \m} & =0 \ .}}
This is solved by taking
\eqn{aptxx}{
\eqalign{
{\hat \l} & = {\xi \over \sqrt{1+ \cos \f}}
\cr
{\hat \m} & = {\zeta \over \sqrt{1+ \cos \theta}}}}
for constant $\xi$, $\zeta \in \bR$ and $B$, $\f$ and
$\theta$ are determined by
\eqn{apuxx}{
\eqalign{
\pd{\bar w} (2 \f +i (K+2 \log {B \over A})) & =
2i K_i \pd{\bar w} z^i
\cr
\pd{w} (-2 \theta -i (K+2 \log {B \over A})) & =
-2i K_i \pd{w} z^i}}
We note that these solutions generically preserve
$1 \over 2$ of the supersymmetry.

It is straightforward to check that these conditions
ensure that the
 scalar and Einstein field equations hold.

\subsection{Quarter Supersymmetric Solutions}

We shall concentrate on the case $\Gamma^5 \Gamma_{\ubx}
 \Gamma_{\uby} \e =  -\e$
for which $|\t^i|=1$, as
the case $\Gamma^5 \Gamma_{\ubx} \Gamma_{\uby} \e =  +\e$
follows by
analogous reasoning. When
 $\Gamma^5 \Gamma_{\ubx} \Gamma_{\uby} \e =  -\e$,
the Killing spinor can be written as
\eqn{kla}{\eqalign{
\e_1 & = \sin \f^i \pmatrix{ \l^i \cr \l^i}
\cr
\e_2 & = -(1+ \cos \f^i) \pmatrix{ \l^i \cr \l^i}}\ .}
In addition,
\susya\ may be written as
\eqn{klb}{
\eqalign{
(\s^1 +1) \big[ -2i \pd{\bar w}A {\bar \eta} -AB e^{K \over 2}
 {\bar f} \eta \big] =0} \ .}
Then defining
\eqn{klc}{
\eqalign{
\Lambda = 2i \pd{\bar w} A  \big(ABe^{K \over 2} {\bar f}
\big)^{-1}}}
we note that \susya\ is equivalent to
\eqn{kld}{
\eqalign{
(\s^1 +1) (\Lambda {\bar \eta}+ \eta) = 0} \ .}

So, for the case $\Gamma^5 \Gamma_{\ubx} \Gamma_{\uby} \e = - \e$,
\susya\ and \susyc\ imply that $z^i =z^i(w)$ is holomorphic
 iff $\g^{i \bar j} D_{\bar j} {\bar f}=0$ . If however
 $\g^{i \bar j} D_{\bar j} {\bar f} \neq 0$
then $\t^i = \Lambda$ and
\eqn{xcd}{
\Lambda = e^{i \f}}
 for $\f = \f^i \in \bR$.

It is also necessary to examine \susyb\ . This constraint
 may be rewritten as
\eqn{kle}{
4 \pd{\bar w} {\hat \e} -2i  \pd{\bar w} \log {B \over A}
\Gamma_\ubx \Gamma_\uby {\hat \e}
+i \Gamma^5 (-\pd{\bar w}K +2 K_i \pd{\bar w} z^i) {\hat \e}=0 \ .}
where ${\hat \e} = A^{-{1 \over 2}} \e$.
Then  $\t^i = \Lambda$ implies that
\eqn{klf}{
- \pd{\bar w} z^i = A^{-1} \pd{\bar w} A {\bar f}^{-1}
 \g^{i \bar j} D_{\bar j} {\bar f}\ .}
The supersymmetry constraints are solved by
taking
\eqn{klg}{
\eqalign{
{\hat \e}_1 & = \sin \f \pmatrix{{\hat \l} \cr {\hat\l}}
\cr
{\hat \e}_2 & = -(1+\cos \f) \pmatrix{{\hat \l} \cr {\hat \l}}}}
for ${\hat \l} \in \bR$. Hence \susyb\ is equivalent to
\eqn{klh}{
\eqalign{
4 \pd{\bar w} ({\hat \l} \sin \f) +i (1+\cos \f)
(\pd{\bar w} (K+2 \log {B \over A})-2 K_i \pd{\bar w} z^i)
{\hat \l} & =0
\cr
-4 \pd{\bar w}((1+\cos \f) {\hat \l}) +i \sin \f
(\pd{\bar w} (K+2 \log {B \over A})-2 K_i \pd{\bar w} z^i)
{\hat \l} & =0} \ .}
This is solved by taking
\eqn{kli}{
{\hat \l} = {\xi \over \sqrt{1+ \cos \f}}}
for constant $\xi \in \bR$ and $B$ and $\f$ are determined by
\eqn{klj}{
\pd{\bar w} (2 \f +i (K+2 \log {B \over A}))= 2i K_i \pd{\bar w} z^i}
From these expressions, it follows that these solutions preserve
$1 \over 4$ of the supersymmetry.
Again, it is straightforward to check that the conditions \xcd\ ,
\klf\ and \klj\
are sufficient to
ensure that the scalar and Einstein field equations hold.

\section{Half-Supersymmetric $G_2$-Instanton Solutions}

Further simplifications of the supersymmetry constraints may be
obtained in the special case when the $\cN=1$, $D=4$ supergravity
theory arises from the compactification of $M$-theory on a
manifold with holonomy $G_2$.
In particular, as a consequence of the homogeneity properties
of the $G_2$ manifold volume, we have
\eqn{homax}{
\g^{i \bar j} K_{\bar j} =  s^i \ .}

The conditions presented in the previous section imply that
\eqn{kkconsta}{
\pd{w} (\theta - \f) =0 \ ,}
so that $\theta - \f = {\rm \ const}$.
Furthermore, we note that
$\pd{\bar w} A e^{i \theta} + \pd{w} A e^{i \f}=0$,
so it follows that $A$ depends only on some linear
 combination of $x$ and $y$.
Without loss of generality we shall take $A=A(x)$. Then
\apqxx\ implies that
$z^i = z^i (x)$, and hence all fields depend only on $x$.
 This then fixes the constant to be
\eqn{klconst}{
\theta - \f  = (2n+1)\pi \ ,}
for $n \in \bZ$.
In addition, we require that
\eqn{pcndx}{
\pd{x} \big( 2 \f +i (K + 2 \log {B \over A}) \big) =
 2i K_i \pd{x} z^i\ ,}
\eqn{hombx}{
\pd{x} z^i = -A^{-1} \pd{x} A \big(s^i + {\bar f}^{-1}
\g^{i \bar j} \pd{\bar j}
{\bar f} \big) \ ,}
and
\eqn{homcx}{
AB e^{K \over 2} \bar f e^{i \f} = i \pd{x} A\ .}
Now, \pcndx\ implies that $B=A$ and
\eqn{auxav}{
2 \pd{x} \f = K_i \pd{x} p^i, }
where we recall for the $G_2$ ansatz under
 consideration here, $K_i \in \bR$.
It is straightforward to see that contracting
the imaginary portion of
\hombx\ with $K_i$ one obtains the relation
\eqn{auxbv}{
{1 \over 2} K_i \pd{x} p^i - \pd{x} \big({\rm Arg}
\ f \big)=0 \ ,}
and so \auxav\ and \homcx\ are consistent, as we expect.
We shall set $\f = {\rm Arg} f +{\pi \over 2}$
so that
\eqn{conseqa}{
A^{-2} \pd{x} A = e^{K \over 2} |f|\ ,}
and we observe that a discontinuity jump of $\pi$ in $\f$
induces a sign change
in \conseqa\ .

\subsection{Instantonic Solutions}

We proceed to solve the half-supersymmetric
constraints \hombx\ and \homcx\
using the instanton induced superpotentials
\supp\ and \sumsup\ .
Taking
first the single instanton cover \supp\ we note
 that \hombx\ implies
that $\pd{x} p^i=0$. We solve this by setting without loss of
generality $p^i=0$ and as with this
choice $f \in \bR$ we take
$\f={\pi \over 2}$ ($\theta = -{\pi \over 2}$).
In addition, \hombx\ may be rewritten as
\eqn{kcond}{
K_i = A^{-2} \x_i - \d_i}
for real constants $\x_i$ and \homcx\ is
\eqn{difch}{
f e^{K \over 2} = - \pd{x} A^{-1} \ .}
We note that in the case when $\x_i=0$ for
 $i=1, \dots , b_3$, $K_i$, $s^i$ and
$\T$ are constants and $A = \zeta x^{-1}$ for constant
$\zeta$, so the spacetime
is simply anti-de-Sitter.
For $\x_i \neq 0$, it is clear
 from  \kcond\ that at supersymmetric extrema
for which $D_i f=0$, we  require $A \rightarrow \infty$.

 A more interesting solution is obtained if we take
$\d_i = \a \x_i \neq 0$ for constant $\a$. Then
 the solution
obtained is a scaling solution with
\eqn{scal}{
s^i = (A^{-2} - \a)^{-1} \x^i}
for constants $\x^i$ satisfying $\x^i \x_i = 1$.
 Then the K\"ahler potential,
and the scalar potential and its derivatives
are given in terms of $A$ by
\eqn{kapex}{
\eqalign{
K & = \log |A^{-2} - \a| + \s_1
\cr
V & =  \s^2 |A^{-2} - \a| e^{-\a (A^{-2}-\a)^{-1}}
\big[ {A^{-4} \over (A^{-2}-\a)^2} -3 \big]
\cr
{\partial V \over \partial z^\ell} & = \a^{-1}
\s^2 A^{-2} |A^{-2} - \a| e^{-\a (A^{-2}-\a)^{-1}}
\big[ {\a^2 \over (A^{-2} - \a)^2} -2 \big] \d_\ell\ ,}}
for constants $\s_1$, $\s= \m^2 e^{{\s_1}}$. The
spacetime dependence is fixed by \difch\ via
\eqn{sptfix}{
\pd{x} A = \s A^2 \sqrt{|A^{-2} - \a|} e^{-{\a \over 2}
(A^{-2}-\a)^{-1}}\ .}

It is not possible to obtain a closed form
for this solution, however
we may examine its generic properties.
We begin with the case $\a>0$.
This solution has two branches; one with
$x_2 < x < x_1$ and $\a^{-{1 \over 2}}<A< \infty$
and the other with
$-\infty < x < \infty$ and $0<A< \a^{-{1 \over 2}}$.

For the first branch, as $x \rightarrow {x_2}{}^+$,
 $A \rightarrow \a^{-{1 \over 2}}$ and $V \rightarrow \infty$.
 As $x$ increases $V$ decreases and $A$ increases. There is
a global minimum of
$V=V_{min} = -2 e^{\sqrt{2}} \a \s^2$ at $x= x_0$, at
 which $A = \sqrt{2+ \sqrt{2}} \a^{-{1 \over 2}}$.
 For $x>x_0$, as $A \rightarrow \infty$, $V$ increases
with $x$ to attain a (local)
maximum value of $V= V_{max} = -3 e \a \s^2$. We
observe that as a consequence
of \sptfix\ , $A \rightarrow \infty$ at
some finite $x=x_1 > x_0$.
Furthermore, setting $x_1=0$ (after perhaps making
some constant translation),
we note that $A \sim {\k \over x}$ as $x \rightarrow 0$
for $\k <0$ constant.
The behaviour of $A$ and $V$ is sketched in Figure 1.
\vskip 0.5cm

\scalebox{0.95}[0.9]{
\includegraphics{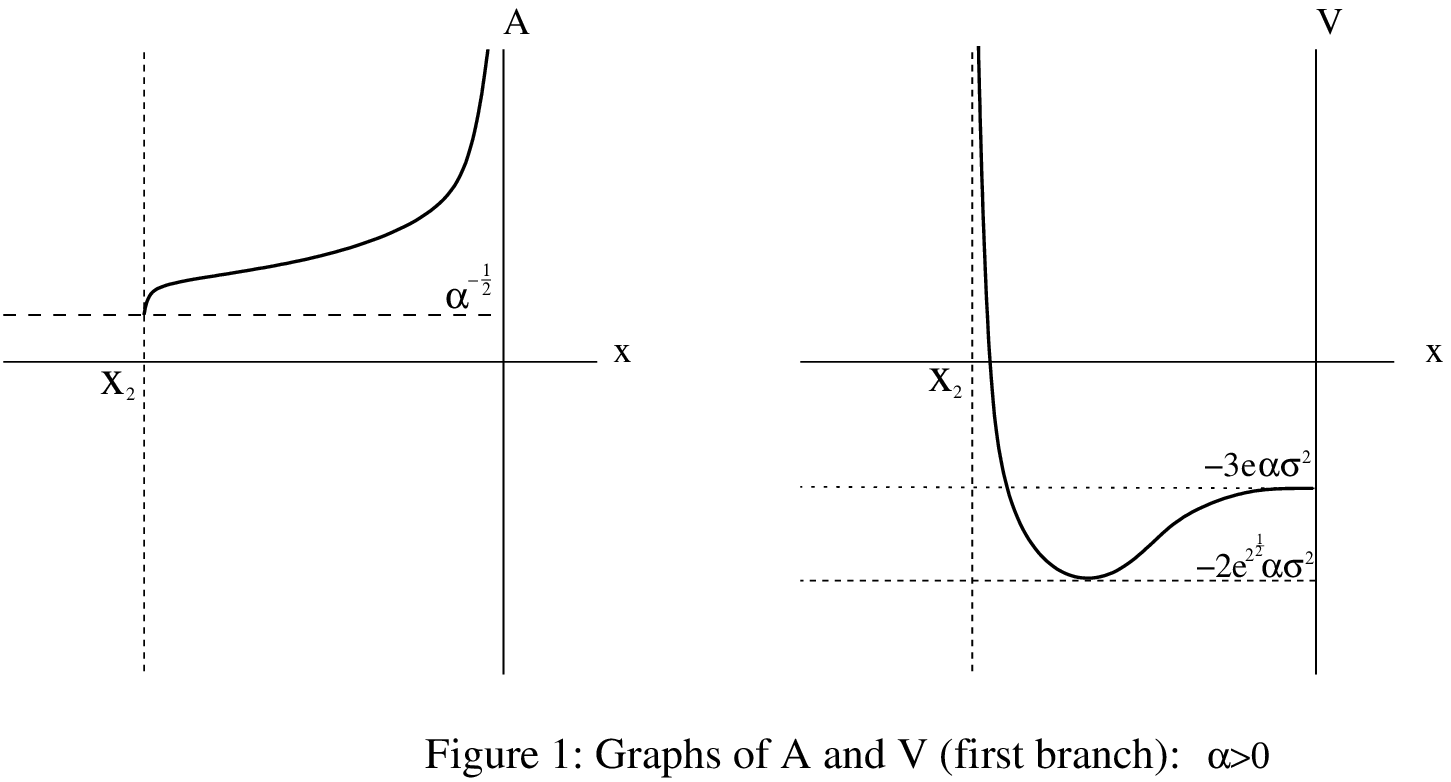}}

It is useful to examine
motion of a test particle in this background. If $\rho$ is
 the affine parameter
along the geodesic then
\eqn{geodesy}{
{d \rho \over dA} = {1 \over \s \sqrt{|(\e A^2 + E^2)(A^{-2} - \a)|}}
 e^{{\a \over 2}(A^{-2}- \a)^{-1}} \ ,}
where $\e = 0$ or $\e = -1$ according to whether we
consider a photon or
massive particle and $E$ is the constant energy. From this we observe
that the particle reaches both $x=0$ and $x= x_2$ in a finite affine
parameter, say $\rho = 0$ and $\rho = \rho_2$ respectively.
As $\rho \rightarrow \rho_2$ the scalars $s^i$, and
 the potential $V$ and all of
their derivatives are unbounded, and the $G_2$
 manifold has a singularity as the $G_2$ volume $\Theta$
 vanishes. As $\rho \rightarrow 0$ the geometry becomes
anti-de-Sitter and all derivatives of $V$ vanish.
 From this it is apparent
that the unique (partial) smooth extension of the
 solution through $\rho = 0$ gives a solution defined
 on a finite interval of affine parameter $\rho$ with a
reflection symmetric double-well potential (Figure 2).

\vskip 0.5cm

\scalebox{0.85}[0.85]{
\includegraphics{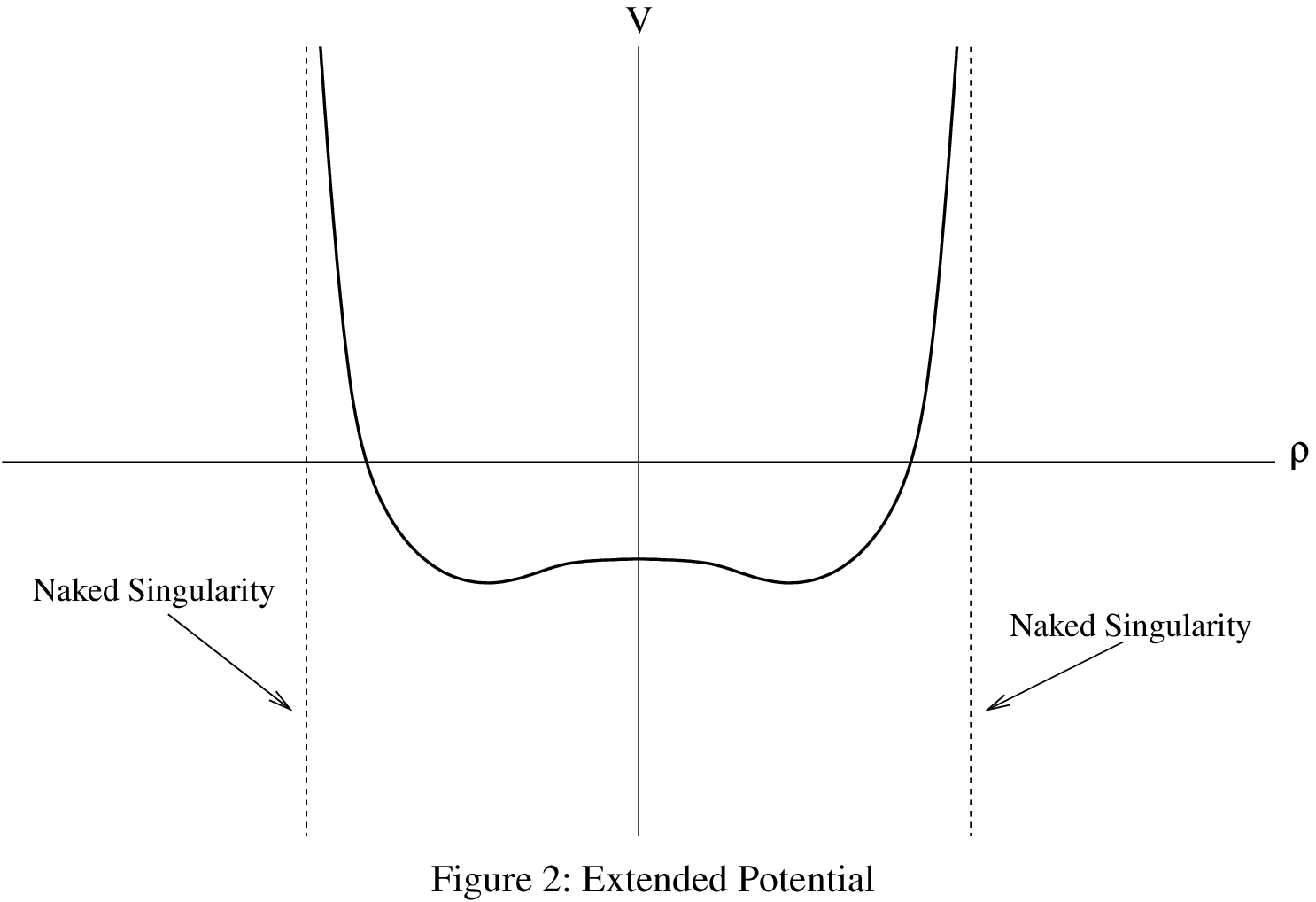}}

At both of the boundaries of this interval
 where $V \rightarrow \infty$, there is
a naked curvature singularity at which
$R \rightarrow \infty$. Although
the potential has two minima, they are not
 supersymmetric. The only supersymmetric
(AdS) extremum is the local maximum at $x=0$.

For the second branch of the $\a>0$ solution,
 $A \rightarrow 0$ as $x \rightarrow - \infty$
 and $A \rightarrow \a^{-{1 \over 2}}$ as $x \rightarrow \infty$.
$V$ increases from $- \infty$ to a global maximum
 of $V_{max} = 2 \s^2 \a e^{-\sqrt{2}}$ at
 $A=\a^{-{1 \over 2}} \sqrt{2-\sqrt{2}}$ and
 then decreases to $0$ as $A \rightarrow \a^{-{1 \over 2}}$.
 The global maximum
is not supersymmetric, nor is the Minkowski minimum;
indeed at the minimum the scalars become unbounded and the $G_2$
volume tends to infinity. Furthermore, an
 analysis of the test particle motion
shows that $A=0$ is reached in finite affine
 parameter. This is a naked
singularity. The behaviour of $A$ and $V$ for
this branch of the solution is
sketched in Figure 3.

\scalebox{0.9}[0.9]{
\includegraphics{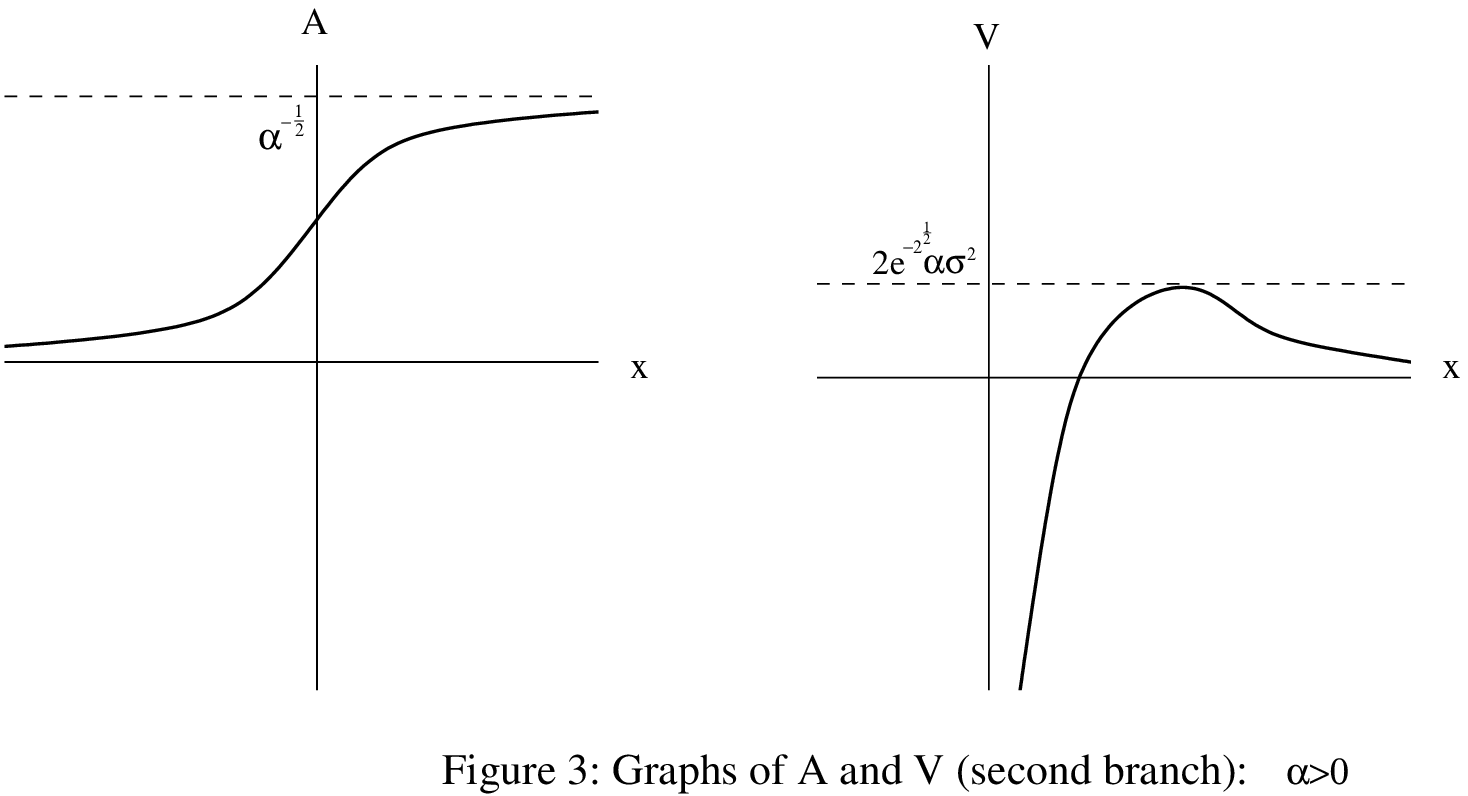}}

For the case of $\a<0$, $A$ is monotonic increasing
 from $0$ as $x \rightarrow
- \infty$ and $A \rightarrow \infty$ as $x \rightarrow 0$
as $A \sim {\k \over x}$ for $\k<0$ constant. $A=0$ is a
naked curvature and $G_2$ singularity.
Just as in the first branch of the $\a>0$ solution, this
solution may be smoothly extended through $x=0$ to give
 a reflection symmetric
potential bounded by two disconnected naked singularities.
 In this case, however,
$V$ has only one extremum, a global supersymmetric
 maximum at $x=0$ of
$V_{max} = 2 \s^2 \a e$. The behaviour of this
solution is sketched in Figure 4.

\scalebox{0.9}[0.9]{
\includegraphics{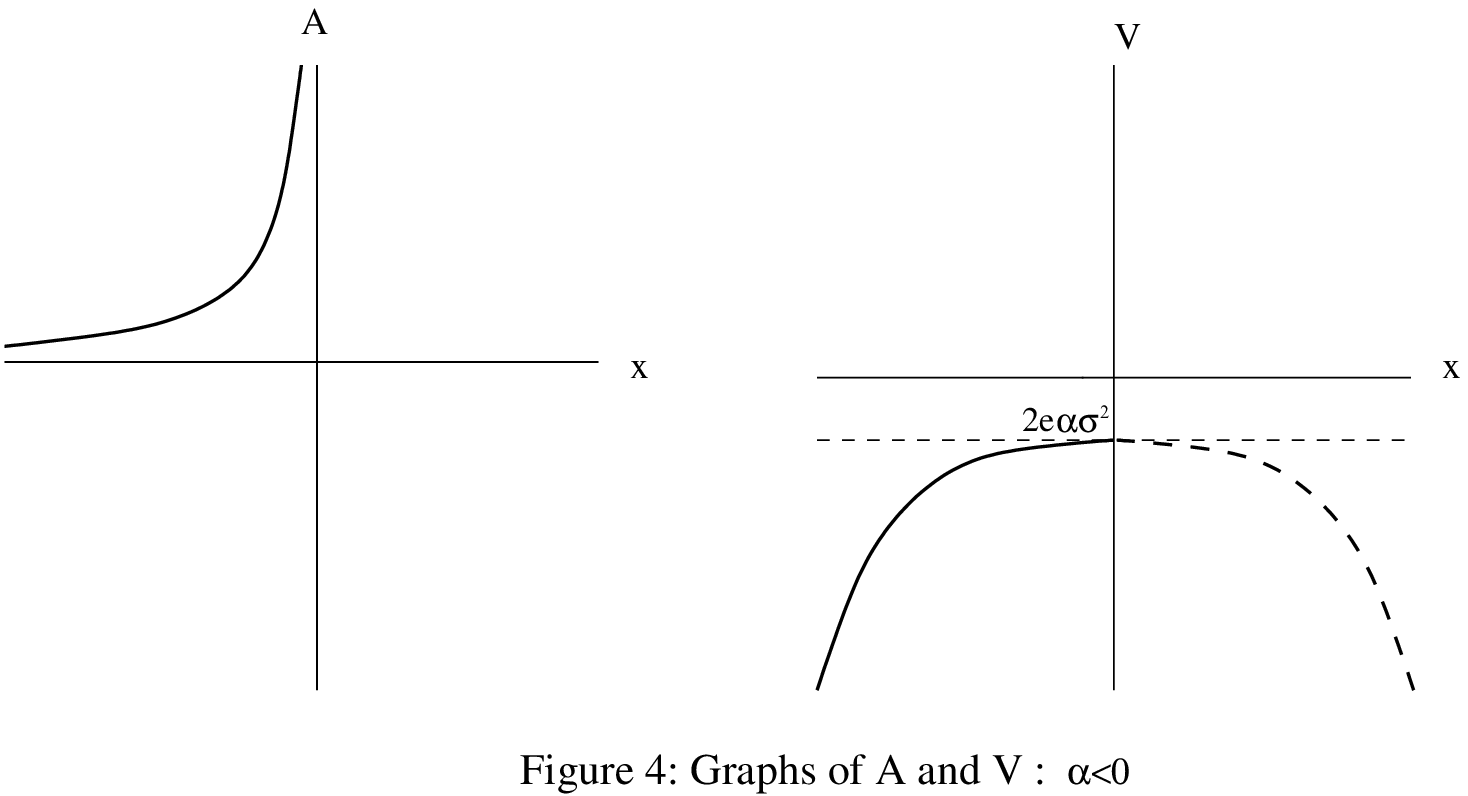}}

For more general solutions where $\x_i$ and $\d_i$
 are linearly independent,
the analysis is considerably more complicated due
to the non-linear structures
on the $G_2$-manifold. However, as a consequence
 of the supersymmetry constraints
it is possible to derive the following useful identities;
\eqn{genident}{
\eqalign{
{R \over 6} & = V + \m^2 e^{K-s^i \d_i}
\cr
{R \over 6} & = \m^2 e^{K - s^\ell \d_\ell}
\big[ \g^{i \bar j} \d_i \d_{\bar j} +2s^j \d_j-2 \big]}}
and from these it is apparent that if
$V \rightarrow \infty$ then $R \rightarrow
\infty$ and there is a $G_2$ manifold singularity.
It is unclear
if this necessarily implies that $A \rightarrow 0$,
 which is a $G_2$ singularity
as the $K_i$ become unbounded; for linearly
independent $\x_i$ and $\d_i$ it
is no longer possible for a $G_2$ singularity
to arise from all of the $K_i$ vanishing.

We can also examine the solutions obtained when one takes the sum of
multiple M2-brane instanton contributions to the superpotential
given by \sumsup\ . In this case, the equations are
 considerably more complicated.
We define
\eqn{cpeq}{
\cP (z) = \sum_{i=1}^\infty {e^{nz} \over n^2}}
so that $f = \m \cP$.
In particular, from this form of $f$ it is apparent that $\hombx$ no
longer implies that $\pd{x} p^i=0$. If, however,
we restrict the solution
to $s>0$ then we may again set $p^i=0$ and $\f = {\pi \over 2}$.
The equations simplify even further if we take
 a scaling solution ansatz
\eqn{compscsl}{\eqalign{
s^i & = s(x) \d^i
\cr
K_i & = s^{-1}(x) \d_i}}
where $\d^i$ are real constants such that $\d_i \d^i=1$.
The K\"ahler potential is given by
\eqn{kpt}{e^{K \over 2} = {\tilde{\s}} s^{-{1 \over 2}}}
for constant ${\tilde{\s}}>0$ , and the
supersymmetry constraints are solved
by imposing
\eqn{fstconstr}{\eqalign{
\pd{x} s & = 2 \s A s^{1 \over 2}(\cP-s \log (1-e^{-{s \over 2}}))
\cr
A^{-2} \pd{x} A & = \s \cP s^{-{1 \over 2}}\ ,}}
where $\s = \m \sqrt{{\tilde{\s}}}$.
The potential may be written as
\eqn{poteqvx}{
V = \s^2 s^{-1} \big[ s^2 \log (1-e^{-{s \over 2}})^2 -2s
 (\cP \log(1-e^{-{s \over 2}})) -2 \cP^2
\big]\ .}
A numerical analysis of these equations
 shows that $A$ and $V$ behave
just as in the second branch of the $\a>0$
 solution for the single cover
instanton superpotential.

So, for these scaling solutions with $\iim z^i=0$,
we have shown that the  spacetimes
typically possess naked curvature and $G_2$ singularities.
It is apparent that they do not produce supersymmetric
AdS minima of the potential, and they do not
 as they stand readily
have an interpretation as domain wall solutions.

More interesting scaling solutions may
 be obtained by allowing $\iim z^i$
to vary. In particular, we shall examine
 the behaviour of the multiple
cover superpotential with the scaling solution
\eqn{compscslb}{\eqalign{
s^i & = s(x) \d^i
\cr
p^i & = p(x) \d^i
\cr
K_i & = s^{-1}(x) \d_i}}
where $\d^i$ are real constants such that $\d_i \d^i=1$.
We allow $p(x)$ to vary and allow for $s<0$ by taking the
appropriate analytic continuation of the superpotential.
To be specific, we shall restrict $\cP$ to the principal
branch of the polylogarithm function. The discontinuities
of this function give curvature discontinuities in the
spacetime geometry which in turn give rise to domain wall
solutions. The K\"ahler potential is given by
\eqn{kptb}{e^{K \over 2} = {\tilde{\s}} |s|^{-{1 \over 2}}}
for constant ${\tilde{\s}}>0$.
Setting
\eqn{feqnb}{
\f = {\rm Arg} \ \cP (x)+ {\pi \over 2}}
where $z = -{1 \over 2}(s+ip)$, \hombx\ and \homcx\ imply
\eqn{fstconstbr}{\eqalign{
\pd{x} s & = 2 \s A s |\cP| |s|^{-{1 \over 2}}
(1 -s \rre (\cP^{-1} \log (1-e^z)))
\cr
\pd{x} p & = 2 \s A s^2 |s|^{-{1 \over 2}} \iim
 (\cP^{-1} \log (1-e^z))
\cr
A^{-2} \pd{x} A & = \s |\cP| |s|^{-{1 \over 2}}\ ,}}
where $\s = \m \sqrt{{\tilde{\s}}}$.
The potential may be written as
\eqn{poteqvxb}{
V = \s^2 |s|^{-1} \big[ s^2 \log (1-e^z)^2 -2s
 \rre (\cP \log(1-e^z)) -2 |\cP|^2
\big]\ .}
We have been unable to find an analytic
solution to the equations \fstconstbr\ .
The problem may be simplified somewhat by considering $A$ and $s$
as functions of $p$. Then \fstconstbr\ implies that
\eqn{simplerb}{
{\eqalign{
s {d \over dp} \big( \iim \log \cP \big) & = {1 \over 2}
\cr
A^{-1} {dA \over dp} & = {1 \over 2 s^2 \iim
\big( \cP^{-1} \log (1-e^z) \big)}}}}
Although we have been unable to solve these
equations analytically, they may be solved
numerically. Of particular interest
is a solution with $s_1 < s < s_0$ for $s_0$,
$s_1 <0$, so unlike
the other solutions, there are no $G_2$ singularities. Furthermore,
$A>A_0$ for some $A_0>0$ constant and when $A \rightarrow \infty$
the divergence is according to $A \sim {\k \over x}$
 for finite $x$ and $\k$ constant,
so this portion of the spacetime may be
continued through this co-ordinate singularity
into another copy of itself. This divergence of the conformal factor
corresponds to a supersymmetric minimum of the potential and
$A=A_0$ corresponds to a non-supersymmetric maximum of the potential.
Although $A$ and $\pd{x}A$ are smooth at $A=A_0$
 there is a discontinuity
jump in the curvature and in $\f$ corresponding to a branch cut in
the polylogarithm function $\cP$. The behaviour of this solution for
$-2 \pi <p<2 \pi$ is sketched in Figure 5.

\scalebox{0.9}[0.9]{
\includegraphics{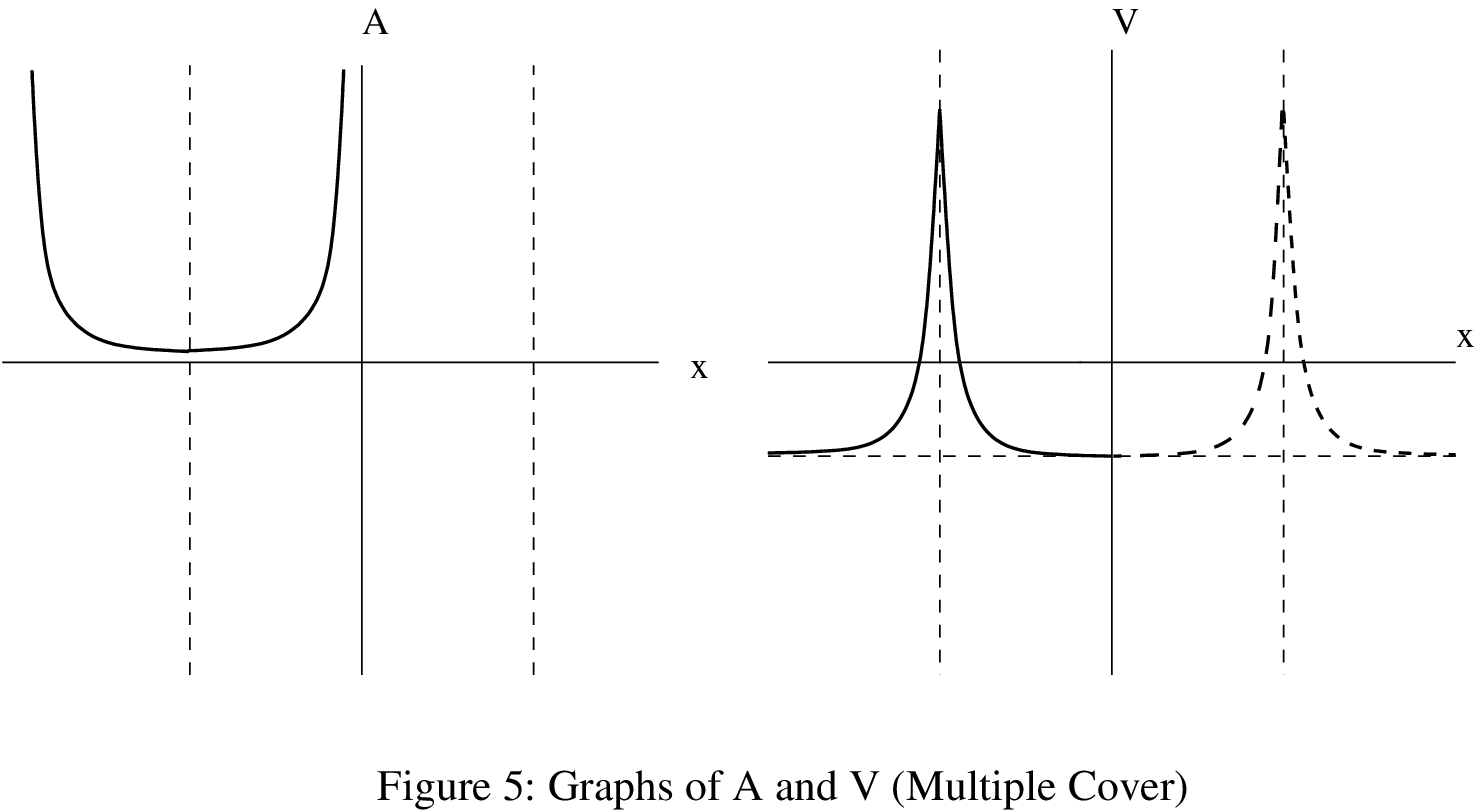}}

We remark that restricting the superpotential
to one particular branch (and so
picking up discontinuities at certain points)
 is essentially equivalent to
truncating the spacetime. Indeed, if we instead
continue smoothly through to
another branch of the polylogarithm, a numerical
 analysis indicates that
just as for the other cases considered previously,
 $s \rightarrow 0$ in a finite
affine parameter, and a curvature and $G_2$ singularity is encountered.

\section{Conclusions}

In this letter we have investigated a class of supergravity solutions
 obtained from
M2-brane instantons wrapping associative
cycles which produce supersymmetric
extrema of the potential. For solutions preserving
half of the supersymmetry arising from
the single cover of M2-brane instantons on an associative
3-cycle, we have demonstrated that
there are no solutions with AdS minima. However, one may, for
example, construct an array of supersymmetric AdS maxima
by truncating by hand the first branch of the $\a>0$ solution
at the minima of the potential and excising the portions of
the spacetime containing the singularities. Gluing on
copies of the portion containing the
 maximum one obtains a solution in which
$V$ is $C^1$ smooth but there is an
 array of curvature discontinuities
at the minima.
For the multiple cover superpotential, the
solution is more promising. In particular, there exists a scaling
solution which {\it does} have supersymmetric AdS minima. It
describes an array of domain walls interpolating between these minima.
However, it is clear that for the solution
 presented here there is no well-defined
finite non-zero charge.

It is apparent that
some aspects of solutions to the supersymmetry constraints presented
in section 3 remain to be addressed. In particular, we have only
been able to construct numerically solutions in which the $G_2$
moduli scale uniformly. More general solutions, even for
the simple superpotential \supp\ satisfying \kcond\
 may have more interesting
spacetime geometries.

In addition, we might expect there to be a
quarter supersymmetric solution which corresponds
 to a superposition
of a stringy cosmic string solution (which has $f=0$
and breaks half of the
supersymmetry) with a domain wall solution
 which will have a more
complicated spacetime geometry than the examples
considered here. Solutions
to the quarter supersymmetric differential
 equations are more difficult to
find. It is perhaps most convenient to eliminate
 the fields $B$ and $\f$;
this then gives
\eqn{secondord}{
{\eqalign{
- \pd{\bar w} z^i = A^{-1} \pd{\bar w} A
 {\bar f}^{-1} \g^{i \bar j} D_{\bar j} {\bar f}
\cr
\pd{w} \pd{\bar w} A = - |f|^{-2} |\pd{w}A|^2 \g^{i \bar j}
 D_i f D_{\bar j} {\bar f}}}\ .}
In principle, given the appropriate boundary
conditions for a string-domain wall
superposition, one may solve these equations numerically.

\vskip 1.0cm

\section*{Acknowledgements}

 I thank G. Papadopoulos for useful conversations.
J.G. is supported
by a EPSRC postdoctoral grant
This work is partially supported by SPG grant PPA/G/S/1998/00613.

\section*{Appendix: Spinor Notation}
It is most convenient to present the
 supersymmetry transformations
in terms of a 4-component Majorana spinor $\e$ with
 real components.
We define  $\sigma^{{{M}}}=(\sigma^{{M}}{}_{\alpha\dot\beta})$
as;
\eqn{apa}{
\eqalign{
\s^{{0}} = \pmatrix{-1 \quad \ \  0 \cr \ 0 \quad -1}
\qquad \qquad
\s^{{1}} = \pmatrix{0 \quad \ \  1 \cr 1 \quad \ \ 0}
\cr
\s^{{2}} = \pmatrix{0 \quad -i \cr i \quad \ \ 0}
\qquad \qquad
\s^{{3}} = \pmatrix{1 \quad \ \ 0 \cr 0 \quad -1}\ .}}

We set $\eta_{\underline{M} \underline{N}}=
 {\rm{diag}} (-1,1,1,1)$,
$\epsilon^{12}=\epsilon_{21}=1$, and to perform
the supersymmetry
calculations we define explicitly
\eqn{apb}{
\eqalign{
\Gamma_{\ubx} = \pmatrix{ 0 \qquad \s^1 \cr \s^1 \qquad 0}
\qquad \qquad
\Gamma_{\uby} = \pmatrix{ -1 \qquad 0 \cr 0 \qquad 1}
\cr
\Gamma_{\underline{0}}= \pmatrix{0 \qquad -i \s^2 \cr -i \s^2
\qquad 0}
\qquad \qquad
\Gamma_{\underline{z}}= \pmatrix{0 \qquad -\s^3 \cr - \s^3
\qquad 0}}}
so that
\eqn{apc}{
\Gamma^5 = \pmatrix{ 0 \qquad 1 \cr -1 \qquad 0} \ .}
With these definitions, the gamma matrices satisfy
the Clifford algebra
\eqn{apdd}{
\Gamma_{\underline{M}} \Gamma_{\underline{N}}
+ \Gamma_{\underline{N}} \Gamma_{\underline{M}}=
2 \eta_{\underline{M} \underline{N}} \ .}

\renewcommand{\baselinestretch}{0.87}
\footnotesize


\end{document}